# Spectroscopic visualization of flat bands in magic-angle twisted monolayer-bilayer graphene: localization-delocalization coexisting electronic states

Ling-Hui Tong[†], Qingjun Tong[†], Li-Zhen Yang, Yue-Ying Zhou, Qilong Wu, Yuan Tian, Li Zhang, Lijie Zhang, Zhihui Qin*, and Long-Jing Yin*

*Key Laboratory for Micro/Nano Optoelectronic Devices of Ministry of Education & Hunan Provincial Key Laboratory of Low-Dimensional Structural Physics and Devices, School of Physics and Electronics, Hunan University, Changsha 410082, China*

[†]These authors contributed equally to this work

\* Corresponding author: zhqin@hnu.edu.cn; yinlj@hnu.edu.cn

**Recent transport studies have demonstrated the great potential of twisted monolayer-bilayer graphene (tMBG) as a new platform to host moiré flat bands with a higher tunability than twisted bilayer graphene (tBG). However, a direct visualization of the flat bands in tMBG and its comparison with the ones in tBG remain unexplored. Here, via fabricating on a single sample with exactly the same twist angle of ~1.13°, we present a direct comparative study between tMBG and tBG using scanning tunneling microscopy/spectroscopy. We observe a sharp density of states peak near the Fermi energy in tunneling spectroscopy, confirming unambiguously the existence of flat electronic bands in tMBG. The bandwidth of this flat-band peak is found to be slightly narrower than that of the tBG, validating previous theoretical predictions. Remarkably, by measuring spatially resolved spectroscopy, combined with continuum model calculation, we show that the flat-band states in tMBG exhibit a unique layer-resolved localization-delocalization coexisting feature, which offers an unprecedented possibility to utilize their cooperation on exploring novel correlation phenomena. Our work provides important microscopic insight of flat-band states for better understanding the emergent physics in graphene moiré systems.**

A tiny twist between two van der Waals atomic layers creates moiré superlattices, which greatly modulate their electronic properties with possible creation of isolated low-energy flat bands [1]. The band flattening boosts the electron interaction effects, opening up new avenues for exploring strongly correlated phenomena. The most celebrated example is the twisted bilayer graphene (tBG), in which the flat-band induced correlated insulating and superconducting states were first observed when the twist angle approaching the so-called magic-angle ~1.1° [2,3]. Band flattening is not limited to tBG, it also appears in twisted multilayer graphene, such as twisted double-bilayer graphene (tDBG) [4-9], with even smaller bandwidth and hence more prominent electron correlation effects [10-13]. Recently, twisted monolayer-bilayer graphene (tMBG, an AB-stacked bilayer twisted with a monolayer graphene) has been predicted as a new flat-band material, which would exhibit much richer tunabilities of the band flattening and topology [14-17]. Subsequently, several transport studies of tMBG have reported the discovery of highly tunable van Hove singularities, correlated states, and Chern insulators [18-21], indicating the versatility of moiré physics harbored in this system. However, a direct visualization of the flat electronic bands of tMBG is still lacking and its microscopic localization knowledge, when compared with tBG, remains elusive, which are indispensable for a full understanding of the rich emergent correlation physics in tMBG.

In this Letter, using scanning tunneling microscopy and spectroscopy (STM and STS) measurements, we report a direct spectroscopic visualization of flat-band states in tMBG near the magic-angle. We also make a comparative microscopic study between tMBG and tBG, which is enabled in our twisted sample with coexisting tMBG and tBG under exactly the same twist angle of ~1.13°. The STS measurement in tMBG shows a clear sharp density of states (DOS) peak at zero-energy, confirming the existence of flat electronic bands. The measured flat-band width in tMBG is slightly smaller than that in tBG, giving a direct experimental evidence of recent theoretical predictions [10,15]. Interestingly, the flat-band states in tMBG exhibit a unique coexistence of localized and delocalized behaviors, i.e., it is localized on the twisted-side while delocalized on the bilayer-side. This intrigue layer-resolved localization-delocalization coexisting nature enables a tunable way to study their interplay on strongly correlated physics. We also perform continuum model calculations which nicely reproduce the observed results.

The twisted multilayer graphene sample is prepared on a highly oriented pyrolytic graphite (HOPG) substrate by surface exfoliation (see the Supplemental Material [22] for details of sample preparation and measurements). Figure 1(a) shows the representative large-area STM topographic image as well as the atomic structure of the studied twisted multilayer graphene. This twisted multilayer consists of a tBG sheet (middle region) partly covered by an extra layer (right region) which is Bernal stacked with respect to the top layer of the tBG, as illustrated in upper panel of Fig. 1(a) (see Fig. S1 [22] for more discussion). Figure 1(b) shows the zoom-in STM topographic images of the tBG and tMBG regions, both of which display the moiré superlattice structures with the same periodicity and orientation. Through measuring the periodicity of the moiré superlattices, we determine the twist angle of ~1.13° for both tBG and tMBG regions. Notice that this angle is very close to the magic-angle for both tBG (~1.1° [2,23]) and tMBG (~1.12° [14]). Furthermore, the corrugation of the moiré superlattices in the tMBG region is much smaller than that in the tBG region under the same measurement parameters (see Fig. S1 [22] for details). These results are quite reasonable because the moiré superlattices of the both regions are formed by a common interlayer twist, but with an AB-stacked top layer placed on the tMBG region, which are consistent with previous STM studies [24-26]. For the tBG, the triangularly arranged bright spots of the moiré patterns are the AA stacking sites, which are connected by the AB/BA-stacked dark regions [23,27-29]. For the tMBG, the bright regions correspond to the ABB sites while the surrounding dark regions are the ABA/ABC-stacked sites [the stacking orders are illustrated in Fig. 1(c)] [18,19]. The topographic image of Fig. 1(a) also reveals a monolayer region, which is a part of the bottom third layer of the tMBG (left region). The obtained atomic-resolution STM image (Fig. S2 [22]) in this region exhibits a clear hexagonal lattice, indicating that the whole twisted graphene sheet is effectively decoupled from the substrate (see Supplementary Material [22] for more discussion) [30,31].

Figure 1(d) shows the representative STS spectra [i.e., the *dI/dV-V* curve reflecting the local density of states (LDOS)] measured in different regions of the graphene sample.

The *dI/dV* spectrum recorded in the tMBG region displays a prominent peak (blue arrow) at the Fermi energy flanked by two smaller side peaks (black arrows). The STS spectrum taken from the bottom monolayer graphene region [inset in Fig. 1(d)] shows that the Dirac point (red arrow) is located closely at the Fermi energy, suggesting a nearly un-doped feature of our sample. Consequently, we attribute the prominent DOS peak at the Fermi energy of the tMBG to the dispersionless flat bands at the charge neutrality point, and the other two smaller peaks to the remote bands in the hole and electron sides respectively [32]. To better understand the tunneling spectra, we perform band structure calculations using continuum model [Figs. 1(e) and 1(f)] [1]. The calculated DOS [Fig. 1(e)] excellently reproduce the spectroscopic features of the tMBG, including the energy locations and relative intensities of the observed three peaks. In the tBG region, we find similar three DOS peaks in the STS spectrum at AA site [inset in Fig. 1(d)], contributing from the flat bands and remote bands in the magic-angle tBG, consistent with previous reports [23,27,28].

The coexistence of tMBG and tBG in a single sample with exactly the same twist angle provides a unique opportunity to directly compare their flat-band electronic properties. We first investigate the bandwidth of the flat bands in these two twisted regions by measuring the full width at half maximum (FWHM) of the flat-band DOS peaks. The measured FWHM of the flat-band peaks in the tMBG region is slightly smaller than the one in the tBG region, which are ~28 ± 2 meV and 34 ± 2 meV respectively (the errors correspond to the standard deviation obtained from dozens of spectra for different AA sites). Such a band narrowing phenomenon also can be clearly distinguished from the normalized STS spectra of the tMBG and tBG [Fig. 1(g)]. We notice that previous theoretical studies [10,15] have predicted that the flat bands in twisted multilayer graphene, including the tMBG and tDBG, have more narrower bandwidths than that in tBG under the same system parameters. However, there is no direct experimental confirmation till now. This is particularly difficult for transport measurement, which usually encounters large variations among different samples [33]. Our spectroscopic measurement therefore provides a direct experimental evidence about this further band narrowing, unambiguously confirming the existence of ultra-flat electronic bands in tMBG.

To further reveal the difference of the electronic structures between tMBG and tBG, we study the spatial dependence of the tunneling spectroscopy. We first focus on the tBG region. Figures 2(a) and 2(b) show the spatially resolved STS contour plot and point spectra for different stacking regions in the tBG moiré superlattices. The low-energy electronic states of tBG exhibit a clearly site-dependent feature. Obviously, the zero-energy flat bands are highly localized at the AA sites, which agrees well with both the single-particle theory [1] and previous STM results [23,27,29,34]. To better reveal the spatial distributions of the electronic states, we measured the energy-fixed STS maps over several moiré supercells. Figures 2(c) and 2(d) show the conductance maps at two energies around the flat-band peak of the tBG. The flat-band states are localized in the AA stacking regions accompanied by the breaking of three-fold ($C_3$) rotational symmetry (see Fig. S3 [22] for more details), which signals the existence of correlation effects [27,29,34]. The STS maps obtained at the electron-side remote band [Fig. 2(e)] and higher energy [Fig. 2(f)] show opposite result: the lower DOS in the AA sites and higher DOS in other regions, also consistent with previous STM report [35]. The excellent agreement between our spectroscopic results and previous studies indicates that this twisted region indeed features as a magic-angle tBG and is electronically decoupled from the substrate.

We now turn to investigate the spatial distributions of the electronic states in the tMBG region. We notice that tMBG has two inequivalent side surfaces, i.e., a twisted-side and a bilayer-side [corresponding to bottom and top surfaces in Fig. 1(a)]. As in tBG region, the twisted-side shows a similar localized behavior of flat bands. This result is obtained from the STS measurements of another tMBG sample whose twisted-side surface is exposed: the flat-band states are strongly localized at the ABB sites (see Fig. S4 [22]). In contrast, the bilayer-side surface, i.e., the exposed surface of the tMBG region in Fig. 1(a), shows distinct behaviors. Particularly, the measured spatially resolved STS spectra on this surface [Figs. 3(a) and 3(b)] exhibit highly homogeneous electronic states throughout the moiré superlattices, especially for the flat-band states: the flat-band DOS peak has almost equal intensity at the three high-symmetry sites. As further evidenced from the STS maps obtained at the flat-band energies [Figs. 3(c) and 3(d)], the flat-band wavefunctions nearly extend over the whole moiré unit cell (see Fig. S5 [22] for more discussion). All of these results suggest a delocalized characteristic of the flat-band states, which is opposite to the observed localized behavior in the tBG region and in the twisted-side of tMBG. The above results indicate that tMBG displays unexpected coexisting localized and delocalized flat-band electronic states. Besides, the high-energy maps measured at the remote band energy [Fig. (e)] and 500 meV [Fig. (f)] also exhibit an entirely different behavior from tBG. The ABB sites have relative higher DOS and one of the ABC/ABA sites shows the lowest DOS, featuring a typical triangular lattice structure. This distinguishing DOS arrangement indicates the breaking of $C_2$ symmetry in the tMBG moiré lattices, which is in contrast to the situation in magic-angle tBG while analogous to that in tDBG [8,9]. Such a broken $C_2$ symmetry is highly related to the distinct band topology of tMBG and leads to the higher tunability of its electronic properties [18-21].

To better understand the experimentally observed intriguing microscopic behavior in the tMBG, we perform a theoretical calculation on its layer- and spatial-resolved LDOS. Figures 4(a)-4(c) show the theoretical LDOS spectra and maps projected on the bilayer-side surface of the 1.13° tMBG, where the electronic states of Fig. 3 are mainly detected. The site-resolved LDOS spectra in Fig. 4(a) nicely reproduces the experimental features in Fig. 3(a), revealing the strong delocalization of the flat-band states. The calculated LDOS map of the flat bands [Fig. 4(b)] also exhibits reasonable correspondence with the measured maps of Figs. 3(c) and 3(d). The discrepancy between theory [Fig. 4(b)] and experiment [Figs. 3(c) and 3(d)] about the detailed positions of the LDOS maxima/minima may come from the approximate treatments of the calculation, which has also been reported in tDBG [8]. Besides, it also shows good agreement between the theoretical high-energy LDOS map [Fig. 4(c)] and the experimental results of Figs. 3(e) and 3(f). For the twisted-side surface of the tMBG [Figs. 4(d)-4(f)], we find that the calculated high-energy LDOS map [Fig. 4(f)] shows similar features in the symmetry and LDOS locations as that obtained on the bilayer-side surface [Fig. 4(c)]. However, the flat bands for the twisted-side display a clearly localized behavior with the electronic states highly concentrate in the ABB stacking regions [shown both in the LDOS spectra and map, Figs. 4(d) and 4(e)], analogous to the scenario of the tBG and also consistent quite well with the experimental result (Fig. S4 [22]). We also calculated the LDOS on the middle-layer of the tMBG, in which nearly the same results are observed as that on the twisted-side surface (Fig. S6 [22]). Our work therefore demonstrates explicitly that the flat bands of tMBG exhibit a unique localization and delocalization coexisting characteristic, i.e., the flat-band electronic states are localized on the twisted-side and delocalized on the bilayer-side.

The localization and delocalization coexisting nature of the flat bands is a distinctive property of tMBG, which provides an unprecedented platform to utilize their cooperation on investigating moiré physics and underlying mechanisms. In tBG, the low-energy flat-band electronic states are highly localized in the AA stacking regions and the Coulomb repulsion between these localized electrons is expected to be responsible for the observed strongly interacting phases [36-38]. While in tDBG, recent theoretical [39] and experimental works [8] have demonstrated the spatially delocalized electronic states and the non-local exchange interaction would play an important role in this case. Our present experimental work therefore indicates that the above two interaction mechanisms may coexist in the tMBG. Indeed, a recent transport study of tMBG has observed the coexisting signature of tBG-like and tDBG-like correlation effects, where a displacement-field-dependent correlation behavior is discovered [18]. This is understandable according to our localization-delocalization coexistence picture. Because the localized states and delocalized states are distributed at different layers of tMBG, a displacement field breaks their degeneracy, making possible a switching between the localization-dominant tBG correlation behavior to delocalization-dominant tDBG correlation behavior.

In summary, the microscopic/spectroscopic characteristics of magic-angle tMBG are studied via STM/STS measurements, accomplished with a direct comparison with tBG. The low-energy flat bands of tMBG and their theoretically predicted narrower bandwidth are directly visualized in the tunneling spectra. Furthermore, we also find that the flat-band electronic states in the magic-angle tMBG shows an interesting coexistence of localized and delocalized behaviors. Our experiment provides a timely knowledge of the microscopic properties of tMBG, which would give important insight for further understanding the novel moiré physics in graphene-based twisted multilayers.

**Acknowledgements**

This work was supported by the National Natural Science Foundation of China (Grant Nos. 12174095, 11804089, 12174096, 51772087, 11904094, 11904095 and 51972106), the Natural Science Foundation of Hunan Province, China (Grant Nos. 2021JJ20026, 2019JJ50034 and 2019JJ50073), and the Strategic Priority Research Program of Chinese Academy of Sciences (Grant No. XDB30000000). The authors acknowledge the financial support from the Fundamental Research Funds for the Central Universities of China.

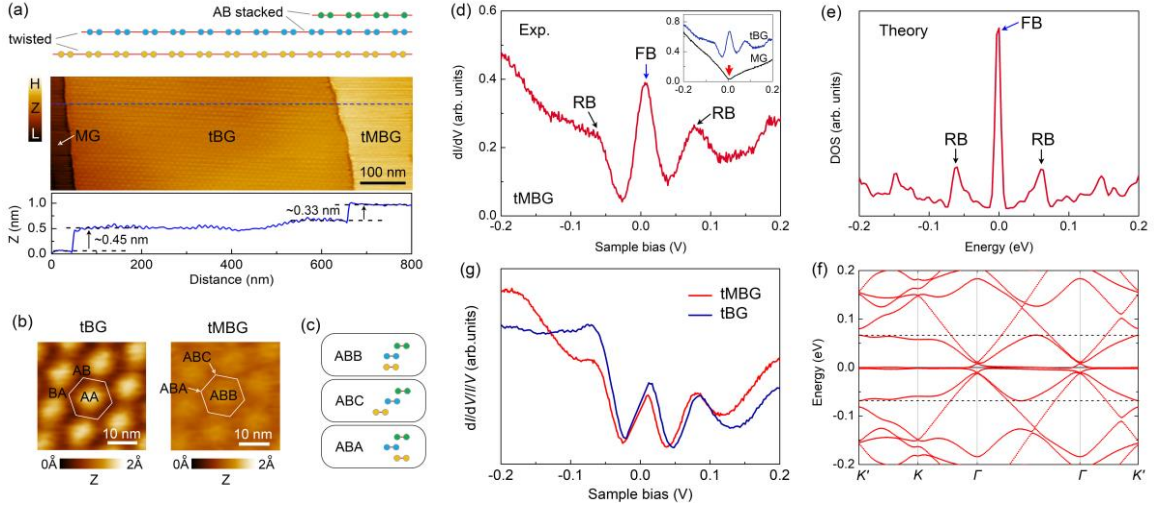

**FIG. 1.** (a) Top panel: schematic of the measured twisted multilayer graphene. Middle panel: large-area STM topographic image (800 nm × 250 nm, $V_b = 0.5$ V, $I = 0.2$ nA) of the coexisting tBG and tMBG on graphite surface. Bottom panel: height profile along the straight dashed line in the middle panel, showing two step structures with the height differences of ~0.45 nm and ~0.33 nm. The top two layers in the tMBG region are AB stacked and together twisted with respect to the bottom monolayer graphene (MG), i.e., the exposed surface of the tMBG is the bilayer-side surface in this sample. (b) Representative zoom-in STM topographic images ($V_b = 0.5$ V, $I = 0.2$ nA) showing the moiré patterns in the tBG and tMBG regions. The white hexagons denote a moiré cell centred at the bright moiré spots. Different stacking configurations are labeled. (c) Side view of three high-symmetry stacking orders in tMBG. (d) A typical $dI/dV$ spectrum of the tMBG obtained at ABB site. Inset shows the STS spectra of the tBG (AA site) and MG regions in (a). FB: flat bands, RB: remote bands. (e) Theoretical DOS of the 1.13° tMBG on the bilayer-side (top) surface. (f) Band structure of the 1.13° tMBG. Dashed lines denote the RB. (g) Normalized tunneling spectra, $dI/dV/I/V$-$V$, of the tMBG and tBG. The curves are averaged from 12 and 21 STS spectra recorded at the ABB and AA sites of different moiré supercells for tMBG and tBG respectively.

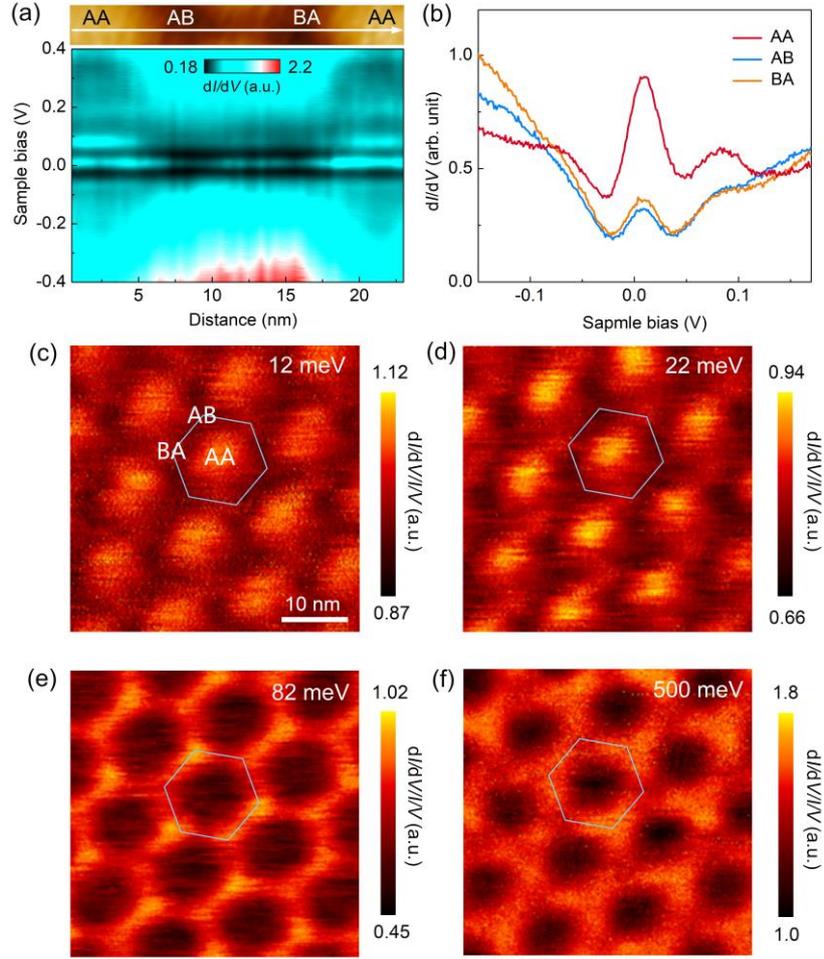

**FIG. 2.** (a) Spatially resolved contour plot of *dI/dV* spectra along the white arrow in the topographic image (upper panel) of the tBG region. (b) Typical *dI/dV* spectra for three different stacking regions (AA, AB, and BA) of the tBG. (c-f) *dI/dV* spatial maps of the same tBG region measured at the energies 12 meV (c), 22 meV (d), 82 meV (e), and 500 meV (f), respectively. *dI/dV* scale bars are normalized by dividing by *I/V* for each panel. Hexagons indicate a moiré supercell centred at AA stacking site.

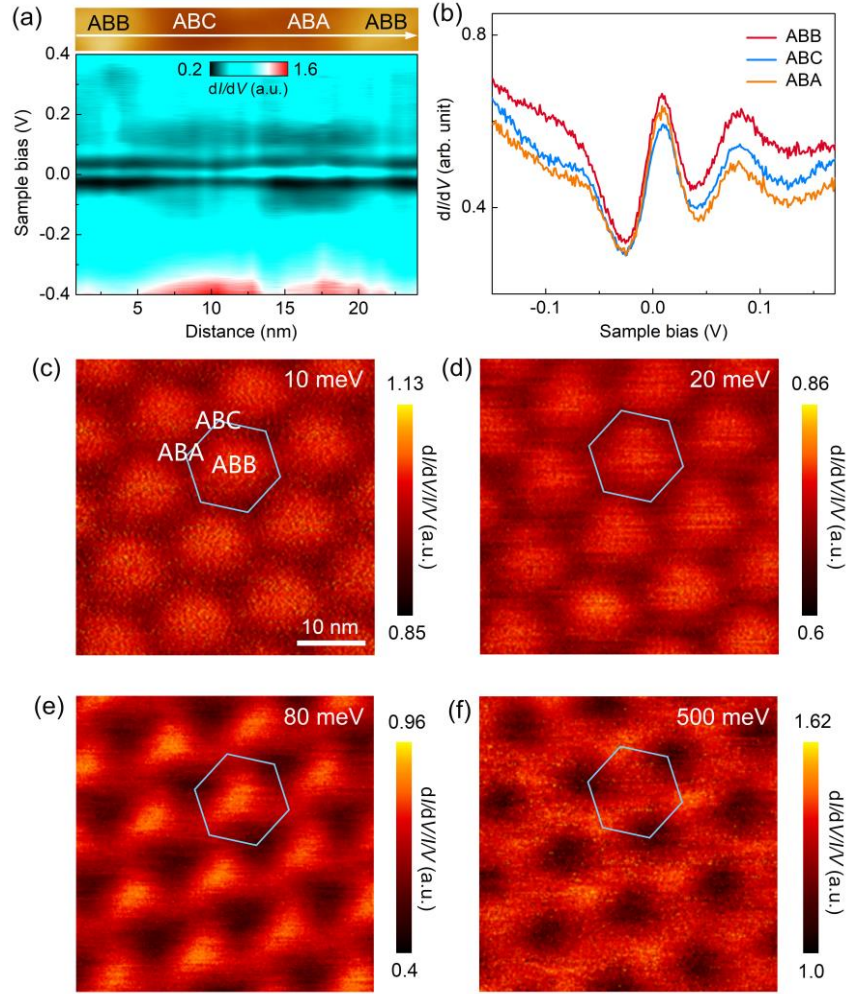

**FIG. 3.** (a) Spatially resolved *dI/dV* spectra along the white arrow in the topographic image (upper panel) of the tMBG region. (b) Typical *dI/dV* spectra for three different sites (ABB, ABC, and ABA) of the tMBG. (c-f) STS conductance maps of the same tMBG region obtained at the energies 10 meV (c), 20 meV (d), 80 meV (e), and 500 meV (f), respectively. *dI/dV* scale bars are normalized by dividing by *I/V* for each panel. Hexagons indicate a moiré unit cell centred at ABB site.

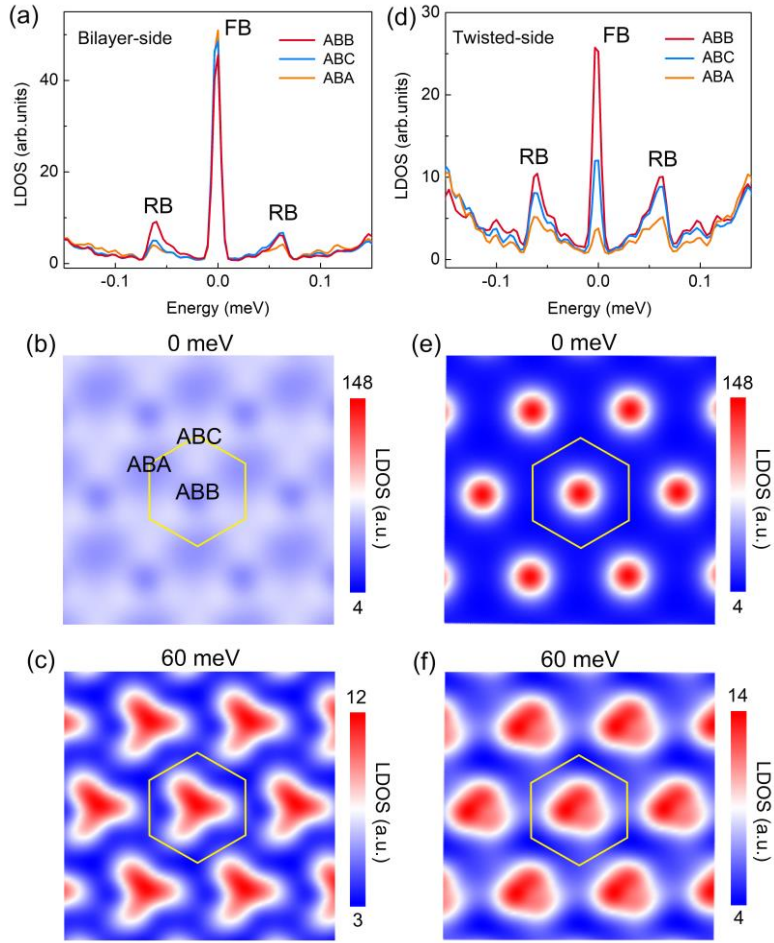

**FIG. 4.** Theoretical LDOS of three high-symmetry sites (a) and LDOS maps for the energies of 0 meV (b) and 60 meV (c) on the bilayer-side surface of a 1.13° tMBG. (d-f) The same as (a-c) but for the twisted-side surface.